\documentclass[a4paper]{spie}  %>>> use this instead for A4 paper
\usepackage{amsmath,amsfonts,amssymb}
\usepackage{graphicx}
\usepackage[colorlinks=true, allcolors=blue]{hyperref}

\title{Impact of water vapor seeing on mid-infrared high-contrast imaging at ELT scale}

\author[a]{Olivier Absil}
\author[a]{Christian Delacroix}
\author[a]{Gilles Orban de Xivry}
\author[a]{Prashant Pathak}
\author[a,\dag]{Matthew~Willson} 
\author[b]{Philippe~Berio} 
\author[c]{Roy~van~Boekel}
\author[b]{Alexis Matter} 
\author[d]{Denis Defrère}
\author[e]{Leo~Burtscher}
\author[f]{Julien Woillez}
\author[e]{Bernhard Brandl}
\affil[a]{STAR Institute, Universit\'e de Li\`ege, All\'ee du Six Ao\^{u}t 19c, B-4000 Li\`ege, Belgium}
\affil[b]{Universit\'e C\^{o}te d’Azur, Observatoire de la C\^{o}te d’Azur, CNRS, Laboratoire Lagrange, Parc Valrose, B\^{a}t H.~Fizeau, 06108 Nice, France}
\affil[c]{Max-Planck-Institut f\"{u}r Astronomie, K\"{o}nigstuhl 17, Heidelberg 69117, Germany}
\affil[d]{Institute of Astronomy, KU Leuven, Celestijnenlaan 200D, 3001, Leuven, Belgium}
\affil[e]{Leiden Observatory, Leiden University, P.O. Box 9513, 2300 RA Leiden, The Netherlands}
\affil[f]{European Southern Observatory, Karl-Schwarzschild-Straße 2, 85748 Garching, Germany}

\authorinfo{E-mail: olivier.absil@uliege.be \\ $^{\dag}$ This work is dedicated to our colleage and friend Matthew Willson, who tragically passed away in January 2022. Matt made strong contributions to the development of focal-plane wavefront sensing for ELT/METIS.}

\begin{document} 
\maketitle

\begin{abstract}
The high-speed variability of the local water vapor content in the Earth atmosphere is a significant contributor to ground-based wavefront quality throughout the infrared domain. Unlike dry air, water vapor is highly chromatic, especially in the mid-infrared. This means that adaptive optics correction in the visible or near-infrared domain does not necessarily ensure a high wavefront quality at longer wavelengths. Here, we use literature measurements of water vapor seeing, and more recent infrared interferometric data from the Very Large Telescope Interferometer (VLTI), to evaluate the wavefront quality that will be delivered to the METIS mid-infrared camera and spectrograph for the Extremely Large Telescope (ELT), operating from 3 to 13~$\mu$m, after single-conjugate adaptive optics correction in the near-infrared. We discuss how the additional wavefront error due to water vapor seeing is expected to dominate the wavefront quality budget at N band (8--13~$\mu$m), and therefore to drive the performance of mid-infrared high-contrast imaging modes at ELT scale. Then we present how the METIS team is planning to mitigate the effect of water vapor seeing using focal-plane wavefront sensing techniques, and show with end-to-end simulations by how much the high-contrast imaging performance can be improved.
\end{abstract}

% Include a list of keywords after the abstract 

\section{INTRODUCTION}
\label{sec:intro}  % \label{} allows reference to this section

\begin{figure}[t]
\begin{center}
\begin{tabular}{cc} %% tabular useful for creating an array of images 
\includegraphics[width=0.5\textwidth]{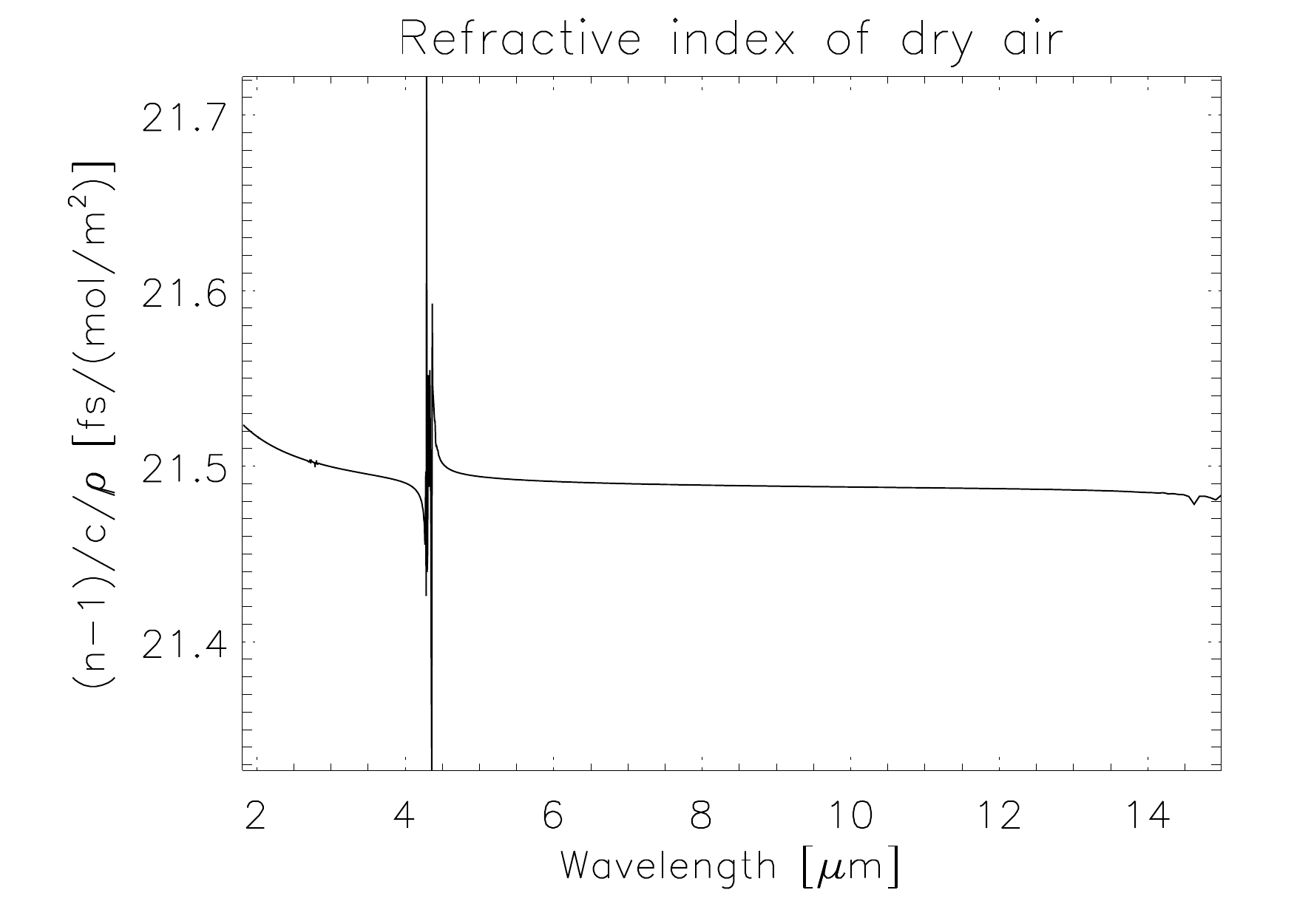} & \includegraphics[width=0.5\textwidth]{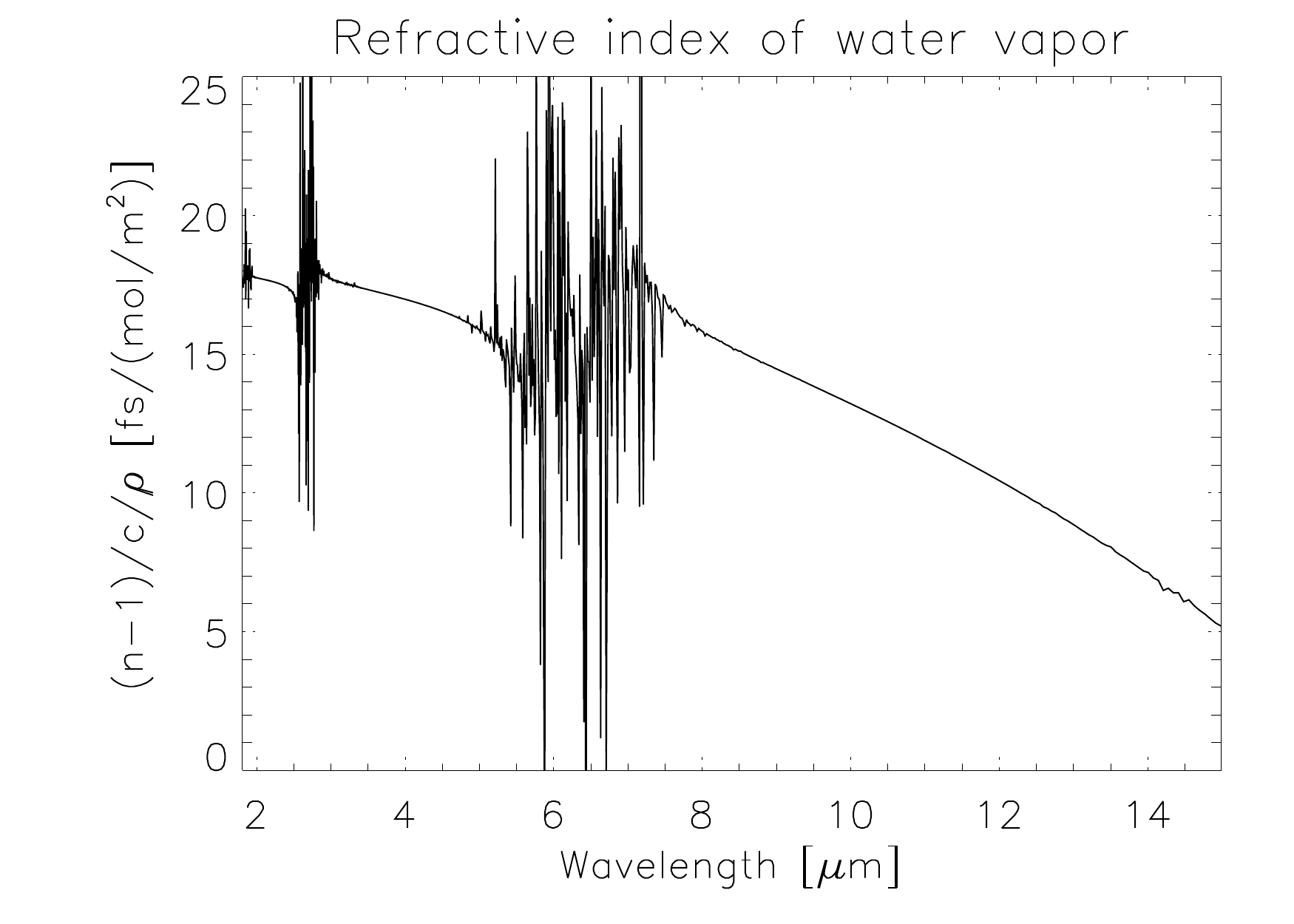}
\end{tabular}
\end{center}
\caption[example]{ \label{fig:refrac_air} 
Reduced refractive indices of dry air (left) and water vapor (right), in units of fs/(mol/m$^2$), based on the model of Mathar\cite{Mathar04}. The effect of CO$_2$ absorption is included in the model of dry air used here, for the sake of illustration.}
\end{figure}

In the standard theory of turbulence, atmospheric seeing results from local variations of refractive index that are transported by the wind. At optical wavelengths, dry air is the dominant source of refractive index variations, although there is a small refractivity (and hence seeing) contribution that can be attributed to atmospheric water vapor (WV), and to other minor species (such as CO$_2$) not considered here. Both dry air and water vapor are slightly dispersive in the visible. At infrared wavelengths, dry air becomes much less dispersive, while WV become increasingly dispersive. This is illustrated in Fig.~\ref{fig:refrac_air}, where we show the reduced refractive indices $\hat{n}(\lambda)$ for dry air and WV, defined as follows:
\begin{equation}
    \hat{n}(\lambda) = \frac{n(\lambda)-1}{c\rho} \, ,
\end{equation}
with $n(\lambda)$ the refractive index, $c$ the speed of light, and $\rho$ the molar density in mol/m$^3$. The reduced refractive index will be expressed in units of femtoseconds per mol/m$^2$ throughout this document.

\begin{figure}[t]
\begin{center}
\begin{tabular}{c} %% tabular useful for creating an array of images 
\includegraphics[width=0.5\textwidth]{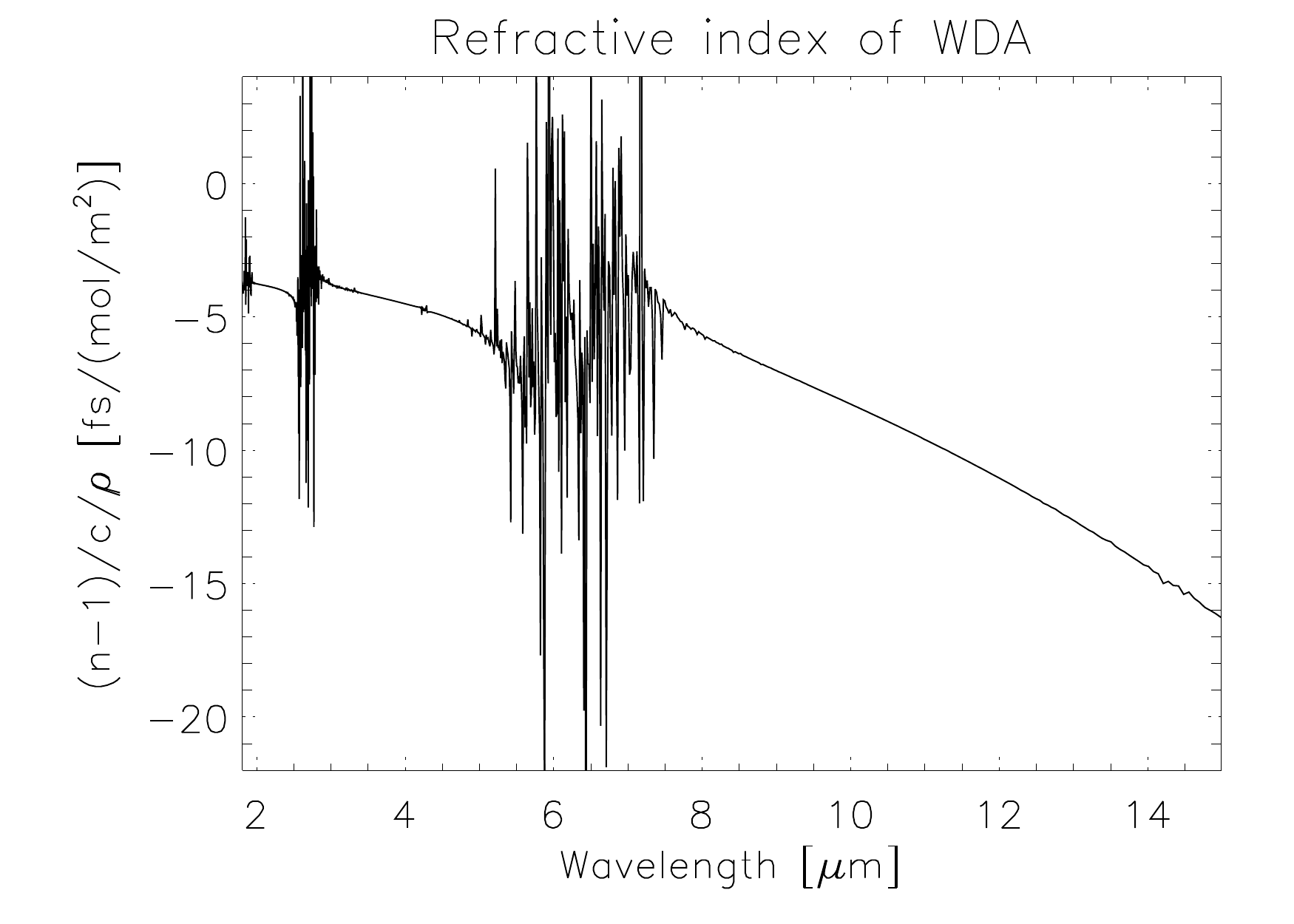}
\end{tabular}
\end{center}
\caption[example]{ \label{fig:refrac_wda} 
Reduced refractive index of WDA, for standard conditions at Cerro Paranal (288~K, 743~mbar).}
\end{figure}

When the wavefront is controlled at optical or near-infrared wavelengths, like for the METIS single-conjugate adaptive optics (SCAO) operating at H--K bands, this dispersion can lead to strong additional wavefront errors at mid-infrared wavelengths, which are not seen by the wavefront control system and therefore not corrected. This effect was extensively studied in the context of mid-infrared interferometers in the early 2000s \cite{Meisner03,Colavita04}, and was first investigated in the context of mid-infrared instrumentation for extremely large telescopes in the late 2000s \cite{Kendrew08}. A convenient quantity to describe the contribution of WV to atmospheric seeing is water displacing air (WDA), introduced in Ref.~\citenum{Meisner03}. Holding the pressure (and temperature) constant, a column density fluctuation of WV induces an opposite variation in the dry air column density. We can thus define a mole of WDA as one mole of WV plus one negative mole of dry air. Consequently, WDA has a reduced refraction index $\hat{n}_{\rm WDA} = \hat{n}_{\rm WV} - \hat{n}_{\rm air}$. At infrared wavelengths, the reduced refraction index of WDA is negative, as shown in Fig.~\ref{fig:refrac_wda}. With the concept of WDA, a given quantity of air can be modeled as an amount of dry air, given solely by the ambient temperature and pressure, plus a quantity of WDA representing the humidity.
 
% ------------------------------------------------------------------------------------------------

\section{WATER VAPOR SEEING} 
\label{sec:wv}

\begin{figure}[t]
\begin{center}
\begin{tabular}{c} %% tabular useful for creating an array of images 
\includegraphics[width=0.5\textwidth]{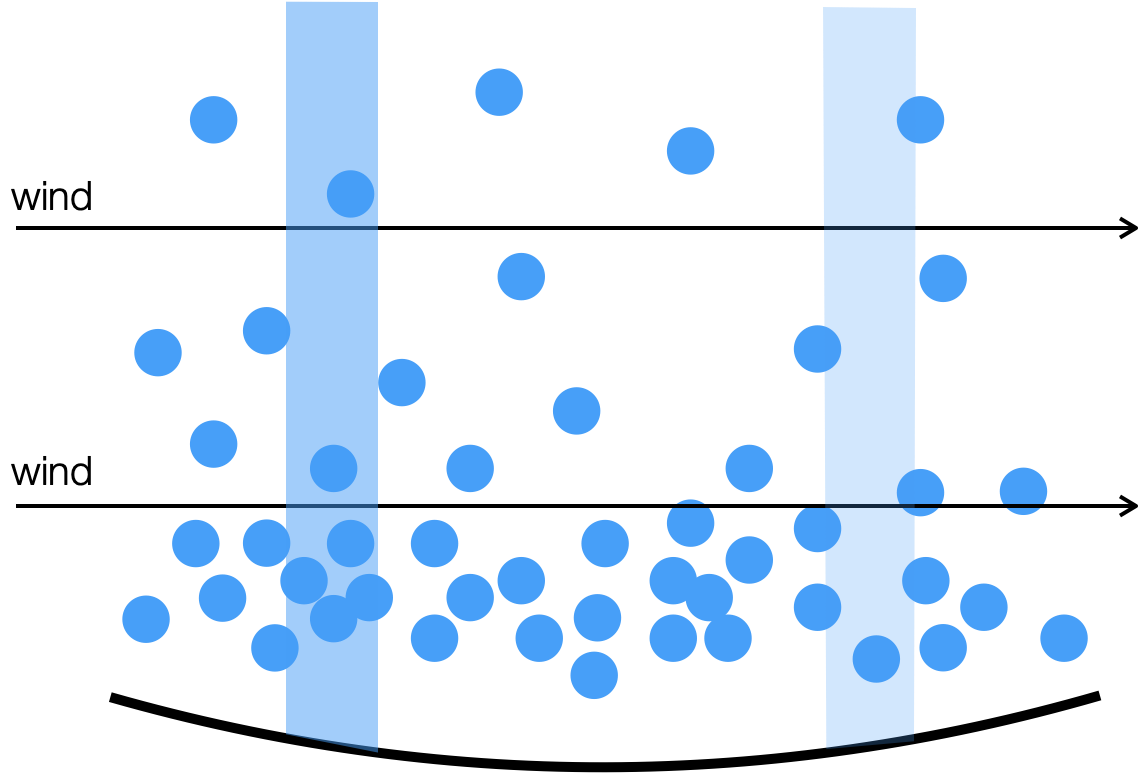}
\end{tabular}
\end{center}
\caption[example]{ \label{fig:wv_seeing} 
Conceptual illustration of WV seeing, where blue circles represent WV molecules in the air, and the rectangular shades of blue the local column density of WV.}
\end{figure}

As can be seen in Fig.~\ref{fig:refrac_wda}, a given amount of WV along the line-of-sight creates a strong chromatic dependence in the mid-infrared refractive index of the atmosphere. The wavefront error seen by the METIS SCAO at H--K bands will therefore not be the same as the wavefront error inside the METIS imager and spectrograph, which cover the L, M and N bands, with an additional delay of about 0.5~fs/(mol/m$^2$) between K and L band, and of 6~fs/(mol/m$^2$) between K and N band. The additional delay (and hence wavefront error) depends on the local column density of WV above the telescope aperture, as conceptually illustrated in Fig.~\ref{fig:wv_seeing}. The non-uniformity of the WV distribution in the air creates additional spatial variations in the wavefront, which are blown over the telescope pupil by the wind, like in the frozen flow turbulence model. It is therefore expected that WV turbulence follows a Kolmogorov – von Karman statistics like dry air turbulence. This hypothesis was verified experimentally in Refs.~\citenum{Masson94} and \citenum{Lay97}.

    \subsection{Measurements from the literature} \label{sub:lit}
    
In order to predict the effect of WV seeing on wavefront errors at mid-infrared wavelengths, one needs to know the spatio-temporal variations of the WV column density. Some measurements of this kind are available in the literature:
\begin{itemize}
    \item Masson\cite{Masson94} used two sub-mm dishes separated by 100~m at Mauna Kea to measure the path length fluctuations due to WV. The measured variance of the column number density is $4 \times 10^{19}$~cm$^{-2}$ over 15 minutes, which corresponds to 0.66~mol/m$^2$ rms. Based on the same data, Colavita et al.~\cite{Colavita04} infer that the median water-vapor seeing from the visible to the L band should be $\sim 1/20$ of dry air seeing, and $\sim 1/7$ at 10~$\mu$m.
    \item A similar setup at the Owens Valley Radio Observatory (OVRO) used by Lay\cite{Lay97} suggests a typical level of fluctuations of 1.5~mol/m$^2$ over a 100-m baseline (see also Ref.~\citenum{Meisner03}). These authors also suggests that the outer scale of turbulence might be significantly larger for WV turbulence than for dry air turbulence, which would lead to generally slower fluctuations than dry air.
    \item Meisner \& Le Poole\cite{Meisner03} mention that group delay measurement with VINCI at VLTI have yielded estimates ranging from 0.5~mol/m$^2$ on a 16-m baseline to about 1.8~mol/m$^2$ on a 66-m baseline, on a 100-s timescale.
    \item Another type of experiment was conducted by Kerber et al.\cite{Kerber15}, using all-sky precipitable water vapor (PWV) measurements with a microwave radiometer at Paranal. These measurements suggest a median variation of 0.07~mm in PWV over the entire sky, down to $27.5\deg$ elevation. That corresponds to 3.8~mol/m$^2$ rms, and should probably be regarded as an upper limit of what could be experienced by single-dish observations along a given line of sight.
\end{itemize}

Based on the results mentioned above, we should expect for a high-quality astronomical observing site like Cerro Armazones to have an average WV column density fluctuation of the order of 1~mol/m$^2$ rms on a 100 m baseline, with a typical range from 0.5 to 1.5~mol/m$^2$ rms. This corresponds to rms column densities ranging typically from $3 \times 10^{19}$~cm$^{-2}$ to $10^{20}$~cm$^{-2}$. The column density variations expressed in mol/m$^2$ can also be converted into a variability in precipitable water vapor (PWV) by using the molar volume of liquid water (18~cm$^3$/mol), leading to a typical range of 9 to 30~$\mu$m of rms PWV.

    \subsection{Measurements at Paranal based on VLTI data} \label{sub:gra4mat}
    
\begin{figure}[t]
\begin{center}
\begin{tabular}{cc} %% tabular useful for creating an array of images 
\includegraphics[width=0.5\textwidth]{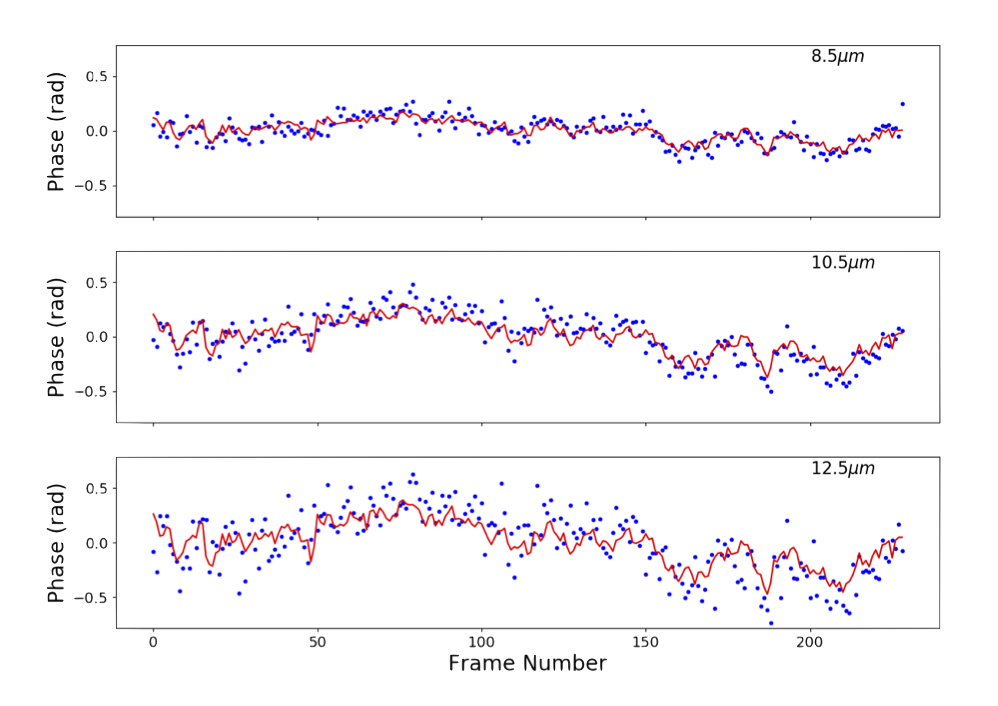} & \includegraphics[width=0.4\textwidth]{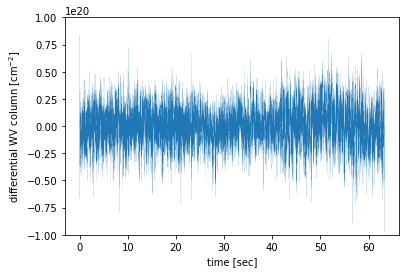}
\end{tabular}
\end{center}
\caption[example]{ \label{fig:gra4mat} 
Left. Comparison of the N-band fringe position estimated by the GRAVITY fringe tracker (red) with the actually measured fringe position with MATISSE (blue), adapted from RD08. Right. Representative time sequence of differential WV column density measured on a $\sim 30$-m baseline at VLTI.}
\end{figure}

While the literature measurements presented above give useful indications on the amount of WV seeing to be expected at the ELT, a more relevant prediction would be based on recent measurements at Cerro Paranal. Such measurements are now available, thanks to the combined operations of the GRAVITY\cite{Gravity17} and MATISSE\cite{Lopez22} instruments at VLTI (an observing mode referred to as GRA4MAT). In particular, the GRAVITY fringe tracker, operating on a few spectral channels within the K band, was used in parallel with MATISSE observations at L or N band. This has allowed to check that a prediction of the mid-infrared fringe position based on a few spectral channels at K band and on a model for the WV refractive index is possible (Berio et al., in prep., see Fig.~\ref{fig:gra4mat}), as already demonstrated more than a decade ago at Keck\cite{Koresko06}. This also provides a direct measurement of the differential WV column density between the four telescopes recombined at the VLTI. In Fig.~\ref{fig:gra4mat}, we show a representative 1-min time series of differential WV column density, which was obtained from the dispersion term computed from the GRAVITY phase delay and group delay estimation for each frame, as described by Koresko\cite{Koresko06}. All GRA4MAT data used here were taken on the smallest baseline configuration of the VLTI, with the Auxiliary Telescopes located at stations A0, B2, C1 and D0, and baselines ranging from about 11 to 36~m.

Based on a series of data sets obtained during the commissioning of the GRA4MAT observing mode, we have been able to measure the rms differential WV column density under various observing conditions. We could not observe a significant dependence of the rms differential WV column density on dry air seeing, while a trend could be identified with respect to the PWV content of the Paranal atmosphere during observations (see Table~\ref{tab:gra4mat}). While this trend is based on a small sample of observations, and does not look linear as expected, it is fair to conclude that the rms differential WV column density increases in high-humidity conditions. From Table~\ref{tab:gra4mat}, we can also infer that the rms differential WV column density measured at Paranal is significantly smaller than the literature measurements described in Sect.~\ref{sub:lit}, with an average of about $2 \times 10^{19}$~cm$^{-2}$ on the short baselines considered here (the shorter baselines being a likely origin for the discrepancy, in addition to the dryness of the Paranal atmosphere). In the rest of this study, we will use this value as our best guess for the typical level of differential WV column density fluctuations at Armazones. This corresponds to a variability of 6~$\mu$m rms in terms of PWV along the line of sight. This quantity can also be converted into a differential delay at a given wavelength with respect to the SCAO operating wavelength, using the refractive index of WDA show in Fig.~\ref{fig:refrac_wda}, and hence into an optical path difference (OPD). The typical OPD fluctuations are of the order of 50~nm at L band, and of the order of 600~nm at N band, which already gives an impression of the level of wavefront error associated with WV seeing.

\begin{table}[t]
\caption{Measured variability of the WV column density at Paranal with GRAVITY and MATISSE.} 
\label{tab:gra4mat}
\begin{center}       
\begin{tabular}{|c|cccc|} 
\hline
\rule[-1ex]{0pt}{3.5ex}  PWV [mm] & 1.4 & 2.0 & 6.6 & 7.0 \\
\hline
\rule[-1ex]{0pt}{3.5ex}  rms WV column density [cm$^{-2}$] & $1.3 \times 10^{19}$ & $1.8 \times 10^{19}$ & $2.2 \times 10^{19}$ & $2.4 \times 10^{19}$ \\
\hline 
\end{tabular}
\end{center}
\end{table}

\begin{figure}[t]
\begin{center}
\begin{tabular}{c} %% tabular useful for creating an array of images 
\includegraphics[width=0.5\textwidth]{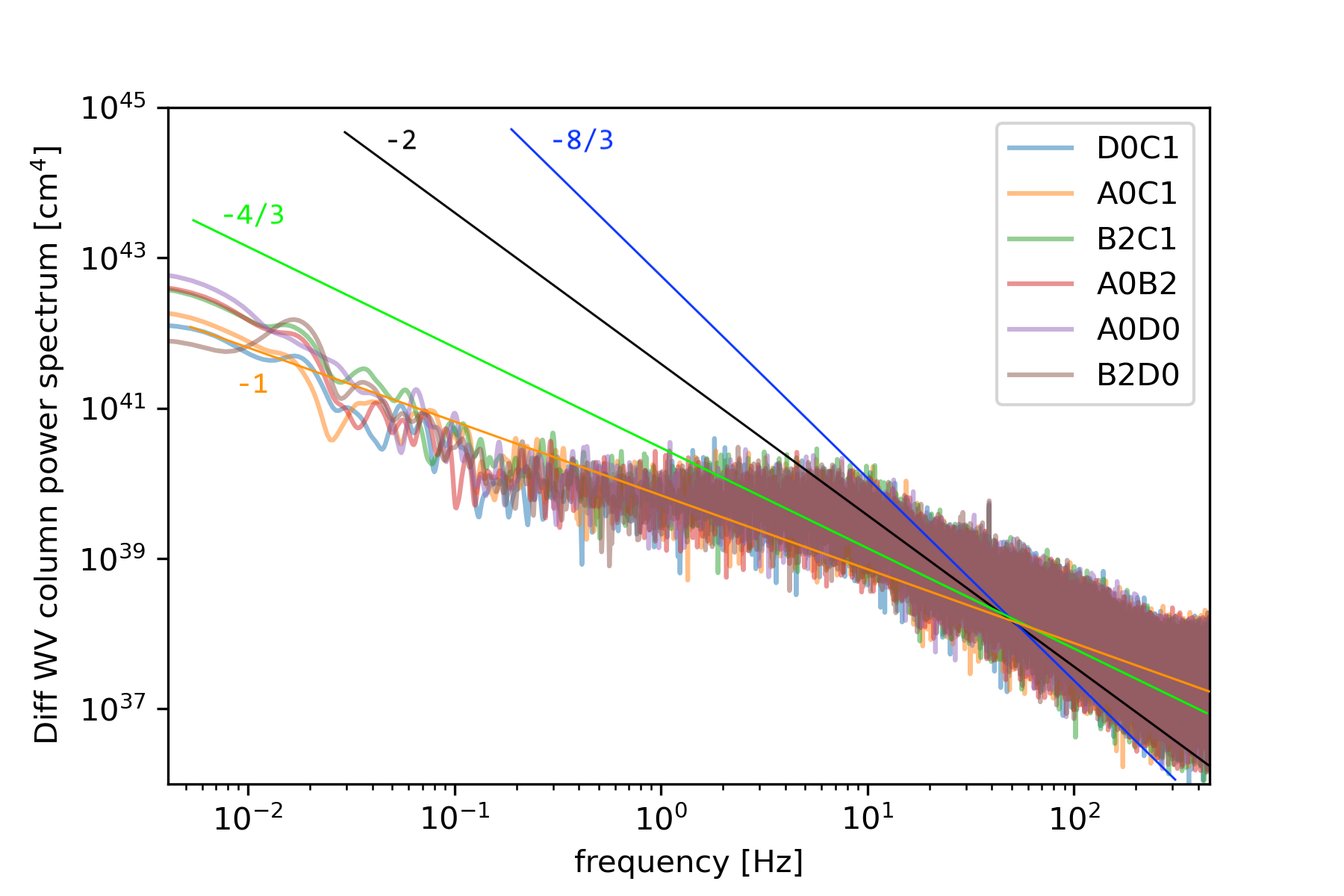}
\end{tabular}
\end{center}
\caption[example]{ \label{fig:psd} 
Measured power spectral density for differential WV column density fluctuations at Paranal, for the six AT-AT baselines used in GRA4MAT observations, obtained in median weather conditions (0.69’’ seeing, 2.0~mm PWV). The PSDs are compared to power laws with various slopes.}
\end{figure}

In addition to measuring the variability of the differential WV column density, we have also checked its power spectral density (PSD) for various observing conditions in terms of dry air seeing and PWV. A series of representative PSDs are displayed in Fig.~\ref{fig:psd}. The expected slope for the PSD of differential WV column density is directly linked to the slope of differential piston, which is equal to $-8/3$ according to the Kolmogorov theory of atmospheric turbulence\cite{Conan95}. This slope is expected to be reduced to $-2/3$ at temporal frequencies below $0.2 v/B$, with $v$ the wind speed and $B$ the baseline\cite{Conan95}, which is of the order of 0.1~Hz in the present case. The overall shape of the observed PSDs does not match these theoretical expectations. The slope at high frequencies ($> 10$~Hz) seems of the order of $-4/3$ instead of $-8/3$, and the apparent knee in the observed PSDs around 10~Hz does not match the expected effect of baseline averaging.
 
A similar shape to the one shown in Fig.~\ref{fig:psd} was observed for the PSD of differential WV column density in four observing sequences obtained under various observing conditions in terms of seeing and PWV, in three different nights. Only in some low-quality data sets with large fringe jumps in the GRAVITY fringe tracker was the observed slope steeper, making it marginally consistent with Kolmogorov predictions. We note that a large variability in the power-law exponent of the phase frequency spectrum was also observed by Masson\cite{Masson94}, with several episodes where the power-law exponent was significantly shallower than the expectation from Kolmogorov theory. The shallow slope of the observed PSDs is not understood at this point, as it would point to a non-physical structure function for WV seeing. In particular, a slope of $-4/3$ for piston would theoretically correspond to a phase structure function that would not depend on the spatial scale, while a shallower PSD slope would be associated with a structure function that decreases with increasing spatial scales. In the absence of a theoretical understanding for the observed PSDs, we will generally assume WV seeing to follow a standard Kolmogorov power spectrum with $-8/3$ slope, as predicted by theory and verified in other observations\cite{Lay97}. But we will also explore the consequences of a shallower PSD on the METIS high-contrast imaging (HCI) performance.

% ------------------------------------------------------------------------------------------------

\section{FROM WATER VAPOR SEEING TO WAVEFRONT ERROR}
\label{sec:wfe}

\begin{figure}[t]
\begin{center}
\begin{tabular}{cc} %% tabular useful for creating an array of images 
\includegraphics[height=5cm]{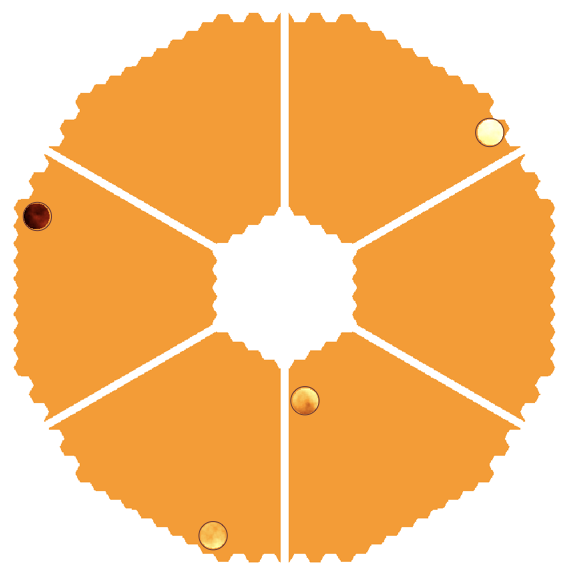} & \includegraphics[height=5cm]{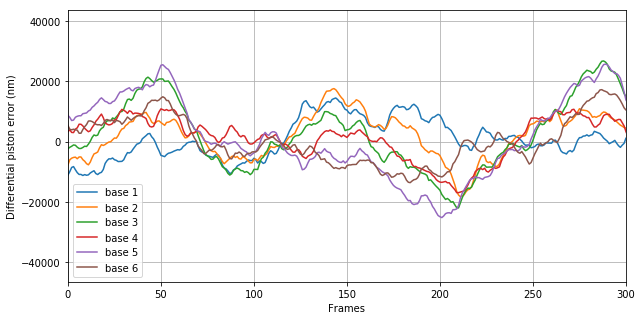}
\end{tabular}
\end{center}
\caption[example]{ \label{fig:scaling} 
Left. Illustration of the considered VLTI baselines superimposed on the ELT-M1 pupil. The color scale shows one realization of the turbulent atmospheric phase on the selected sub-apertures. Right. Corresponding piston measured as a function of time on the six selected baselines. These pistons need to be rescaled to the measured WV-induced phase by GRA4MAT.}
\end{figure}

The differential WV column density measurements performed on the compact VLTI baselines are directly relevant to the ELT, because of the matching spatial scales. Based on the hypothesis that WV seeing follows Kolmogorov statistics, we can then generate a time series of turbulent phase screens on the ELT pupil, and scale this sequence to match the phase variations measured by GRA4MAT on VLTI baselines. In practice, we apply a global scaling factor on the generated turbulent phase screens so that the rms OPD on the six considered VLTI baselines is equal to 50~nm at L band, and 600~nm at N band, which corresponds to a rms differential WV column density of $2 \times 10^{19}$~cm$^{-2}$, as described in the previous section. This strategy is illustrated in Fig.~\ref{fig:scaling}. Here, we used a large outer scale of 500~m, as suggested by literature measurements of WV seeing. Other (smaller) values were also tested for the outer scale, but did not strongly impact the conclusions of this study. The same process was repeated with non-Kolmogorov assumptions, using shallower PSDs for the structure function of refractive index (see Sect.~\ref{sub:non-kolmog}).

The spatial structure of the resulting wavefront has a typical standard deviation of 30~nm rms at L band, and 350~nm rms at N band. Most of the power is in low spatial frequencies, following standard Kolmogorov theory.

% ------------------------------------------------------------------------------------------------

\section{INFLUENCE OF WV SEEING ON METIS HCI PERFORMANCE}
\label{sec:perfo}

    \subsection{Kolmogorov statistics} \label{sub:kolmog}

\begin{figure}[p]
\begin{center}
\begin{tabular}{c} %% tabular useful for creating an array of images 
\includegraphics[width=\textwidth]{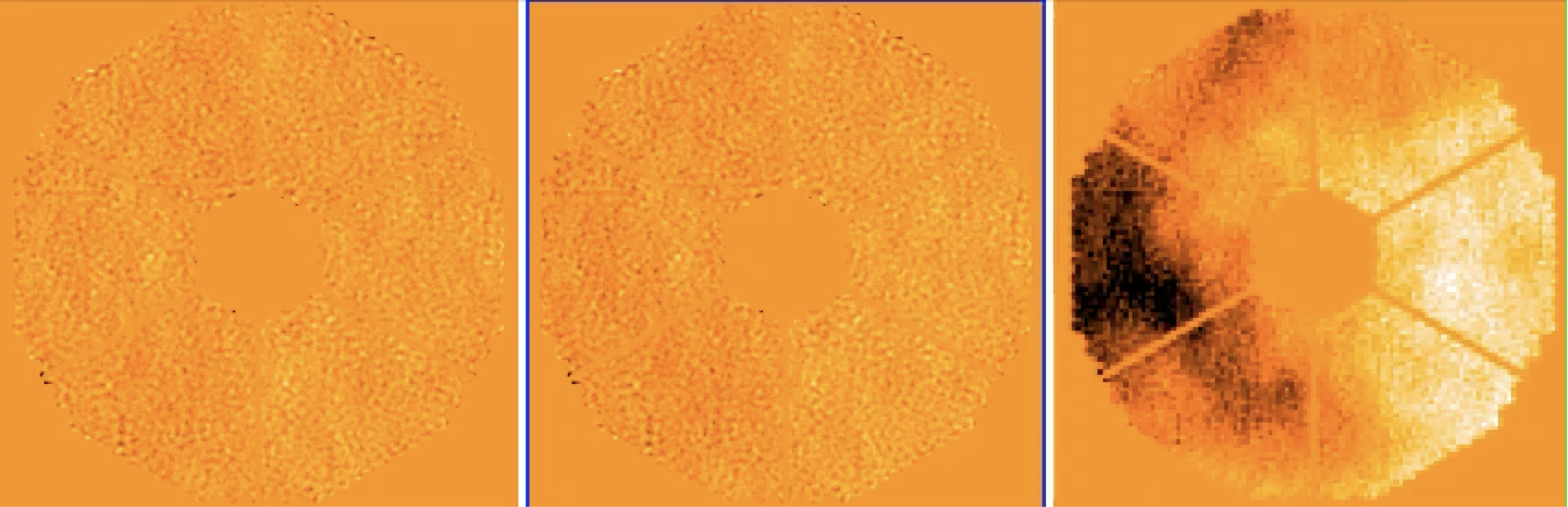}
\end{tabular}
\end{center}
\caption[example]{ \label{fig:scao_wv} 
Representative phase screens after SCAO closed-loop correction (left), and after the addition of typical L-band (middle) and N-band (right) WV-induced wavefront errors.}
\end{figure}

In order to explore the effect of uncorrected WV seeing on the METIS HCI performance, we can simply add the time series of turbulent WV phase screens generated in the previous section to a typical sequence of SCAO residuals produced by COMPASS\cite{Gratadour14}. With a typical residual wavefront error (WFE) of 140~nm in SCAO phase screens, the influence of the additional 30~nm rms of WV-induced WFE is almost negligible at L band, except for the fact that, unlike in residual SCAO phase screens, most of the power is in low spatial frequencies (as expected from Kolmogorov turbulence). Conversely, at N band, the additional 350~nm rms makes a large difference in the amount of WFE seen by the N-band IMG camera of METIS. This is illustrated in Fig.~\ref{fig:scao_wv} for representative phase screens, where the large low-spatial frequency content of WV-induced phase screens is clear.

\begin{figure}[p]
\begin{center}
\begin{tabular}{c} %% tabular useful for creating an array of images 
\includegraphics[width=\textwidth]{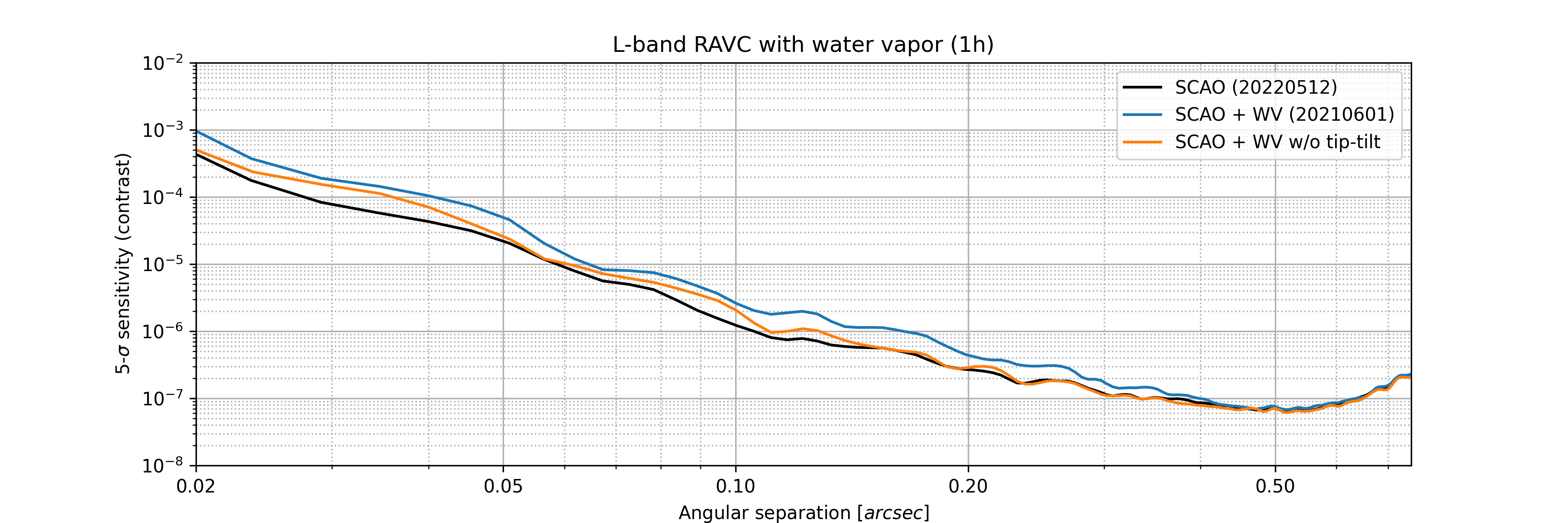} \\
\includegraphics[width=\textwidth]{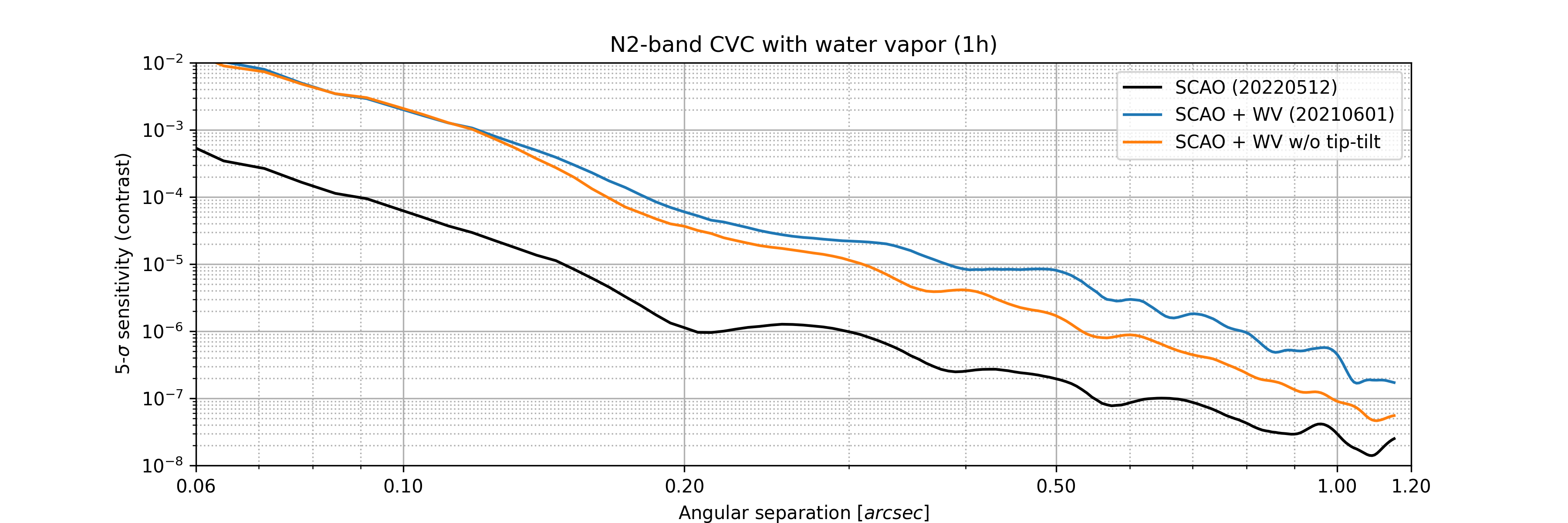}
\end{tabular}
\end{center}
\caption[example]{ \label{fig:perf_kolmo} 
Post-ADI contrast curves for a 1-h observing sequence at L band (top) and N band (bottom), in the presence of SCAO residuals only (black), and of SCAO residual and WV seeing (blue). The orange curves show the performance after completely removing tip-tilt errors from the WV phase screens. These contrast curves were obtained with a ring-apodized vortex coronagraph (RAVC) concept at L band, and a classical vortex coronagraph (CVC) concept at N band, following the METIS design\cite{Carlomagno20}.}
\end{figure} 

The time sequences of phase screens combining SCAO residual and WV turbulence can then be used as an input to end-to-end HCI performance simulations, using the HEEPS simulation package\cite{Carlomagno20}. This is illustrated in Fig.~\ref{fig:perf_kolmo}, where we plot the sensitivity limits in terms of off-axis companions (aka contrast curves) for a simulated 1-h angular differential imaging (ADI) observing sequence, following the same prescription as for the performance simulations described in Ref.~\citenum{Carlomagno20} ($\sim 35\deg$ field rotation in the ADI sequence, and classical ADI processing based on median subtraction, derotation, and mean combination). As expected, the influence of WV seeing on L-band performance is relatively small, although not negligible at small angular separations, due to the relatively large low-spatial frequency content of WV-induced phase screens. Implementing pointing control at a sufficiently fast cadence to follow WV-induced pointing errors (orange curve in Fig.~\ref{fig:perf_kolmo}) would bring the performance back almost to the SCAO-limited performance for most angular separations of interest (beyond $\sim 2.5 \lambda/D$). Conversely, the situation at N band is largely degraded in presence of WV seeing, by up to two orders of magnitudes in terms of achievable contrast. Correcting for pointing errors would only very partly improve upon this situation.

    \subsection{Non-Kolmogorov statistics} \label{sub:non-kolmog}

\begin{figure}[!t]
\begin{center}
\begin{tabular}{c} %% tabular useful for creating an array of images 
\includegraphics[width=\textwidth]{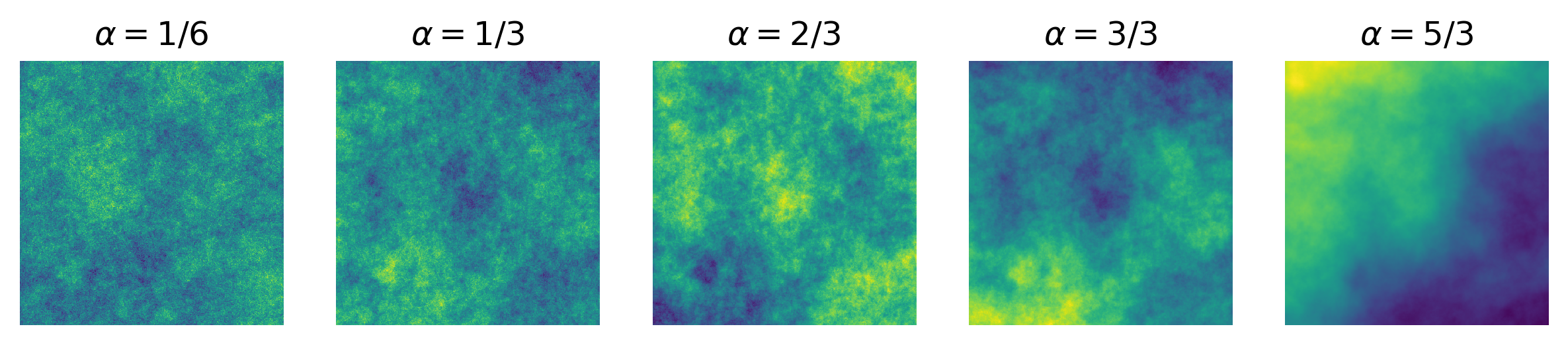}
\end{tabular}
\end{center}
\caption[example]{ \label{fig:struc_funct} 
Representative phase screens obtained with various power-law exponents for the refractive index structure function.}
\end{figure}

In addition to the standard Kolmogorov turbulence spectrum, we tested several custom turbulence power spectra that produce shallower slopes for the PSD of piston. For that, we hacked the COMPASS simulator to change the power-law exponent of the structure function for the refractive index. We tested values ranging from the standard $5/3$ Kolmogorov exponent down to $1/6$ (going down to zero is not possible as the structure function then becomes constant). The spatial structure of representative phase screens is illustrated in Fig.~\ref{fig:struc_funct}, showing that the low-spatial frequency content of turbulence decreases as the power-law exponent of the structure function decreases, while the high-spatial frequency content increases. The phase screens have a slightly larger rms WFE on average.
 
\begin{figure}[p]
\begin{center}
\begin{tabular}{c} %% tabular useful for creating an array of images 
\includegraphics[width=\textwidth]{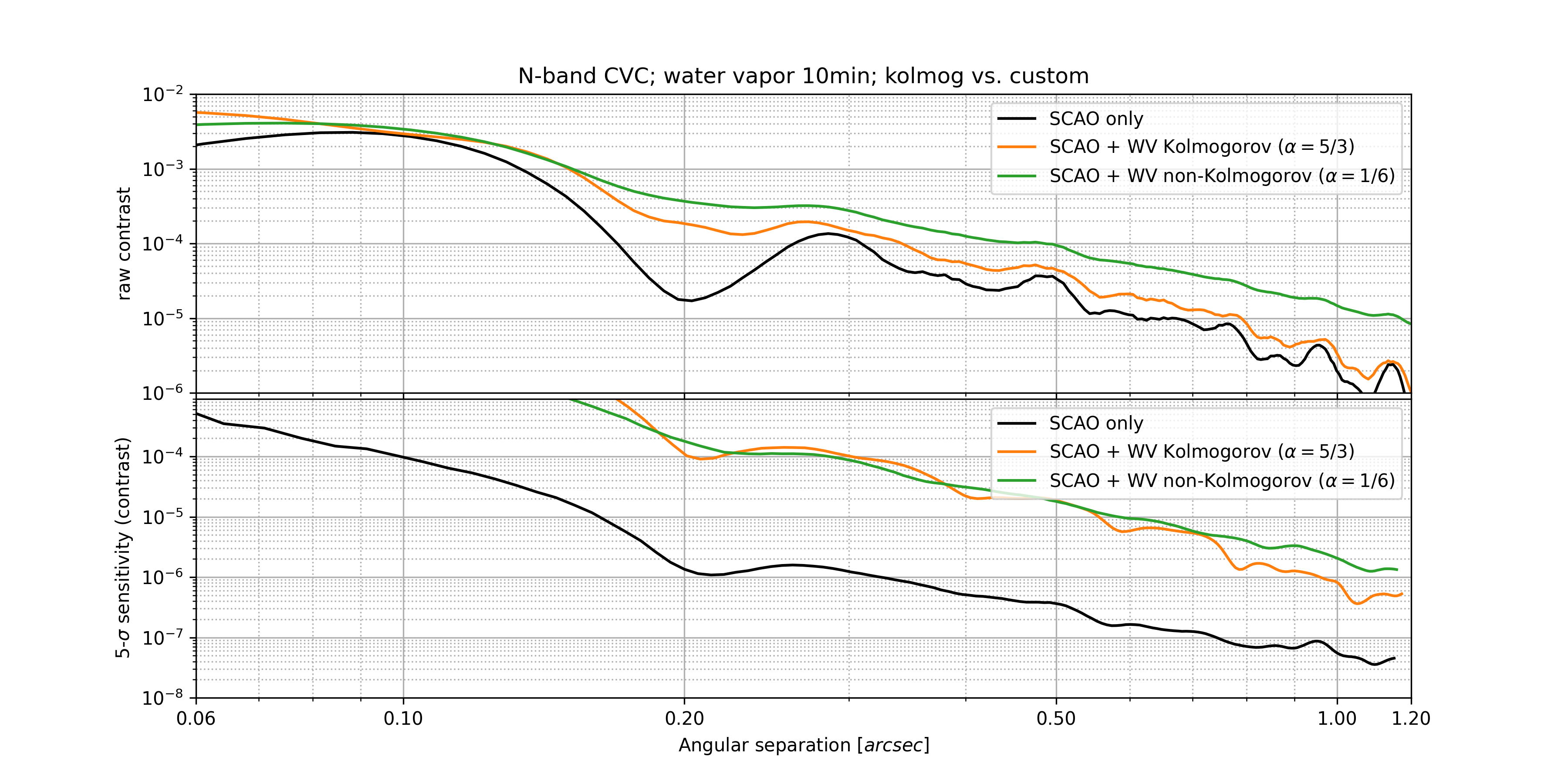}
\end{tabular}
\end{center}
\caption[example]{ \label{fig:perf_nonkolmo} 
Comparison of raw (top) and post-ADI (bottom) contrast curves in the presence of non-Kolmogorov vs Kolmogorov WV turbulence statistics, compared to SCAO-limited performance at N band.}
\end{figure}

We used the lowest power-law exponent of $1/6$ as an extreme case to check how much the HCI performance would be affected by non-Kolmogorov turbulence, using the same simulation strategy as above. The resulting post-ADI contrast curves are illustrated at N band (where the effect of WV is the largest) in Fig.~\ref{fig:perf_nonkolmo}, where it can be seen that the redistribution of aberrations does not significantly change the post-ADI performance. We notice however that the general level of residual starlight making it to the detector is significantly increased in the case of the non-Kolmogorov turbulence, especially at high spatial frequencies. We also anticipate that correcting for tip-tilt only would provide a lower gain than in the Kolmogorov case, because the amount of turbulence power at low spatial frequencies is largely reduced in this case.

While GRA4MAT observations suggest that WV turbulence could deviate from standard Kolmogorov theory, in the following we will generally use Kolmogorov statistics as a more physically motivated assumption. We note however that this might over-estimate the benefits of focal-plane wavefront sensing, which focuses mostly on low-spatial frequencies.

    \subsection{Influence of WV seeing on the METIS Strehl ratio}

\begin{figure}[p]
\begin{center}
\begin{tabular}{c} %% tabular useful for creating an array of images 
\includegraphics[width=0.6\textwidth]{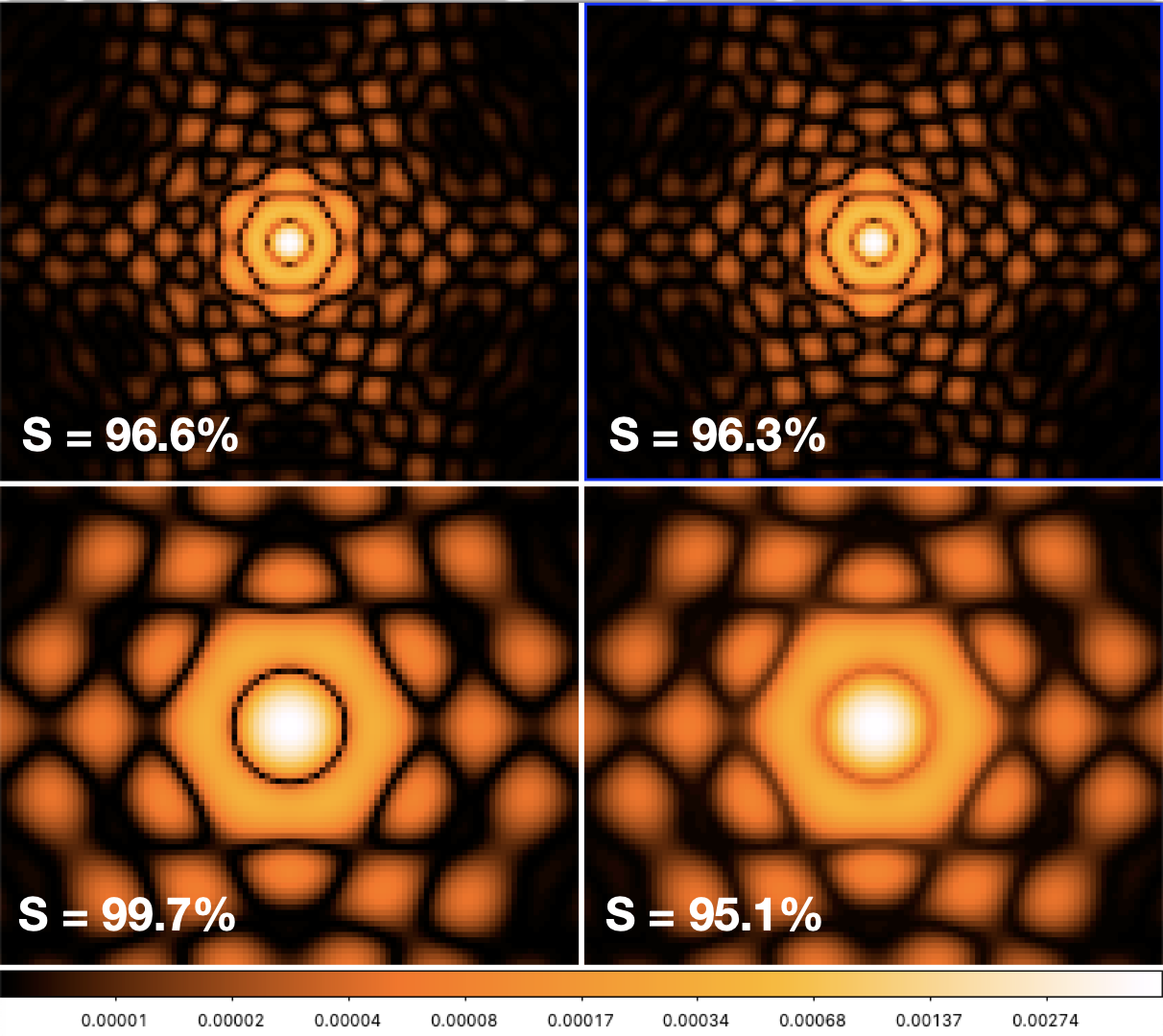}
\end{tabular}
\end{center}
\caption[example]{ \label{fig:strehl} 
Long-exposure PSF of METIS in standard imaging modes for SCAO-limited observations (left) and in the presence of WV seeing on top of SCAO residual (right), respectively at L band (top) and N band (bottom).}
\end{figure}

In addition to exploring the effect of WV seeing on HCI performance, we also explore the effect of WV seeing on the Strehl ratio delivered by standard imaging modes within METIS. This is illustrated in Fig.~\ref{fig:strehl}, where we see a significant degradation in image quality at N band compared to SCAO-limited performance, while L-band imaging is barely affected. The Strehl remains however very high ($>95$\%) in all cases.
 
% ------------------------------------------------------------------------------------------------

\section{MITIGATION STRATEGY}

The large degradation in N-band performance (and to a lesser extent, the mild degradation in L-band performance) due to WV seeing calls for a dedicated mitigation strategy. The METIS instrument, which focuses on HCI applications, includes two focal-plane wavefront sensing (FP-WFS) techniques to measure and correct for aberrations not seen by SCAO. Here, we propose to adapt these two FP-WFS algorithms in order to tackle the effect of WV seeing.

    \subsection{METIS focal-plane wavefront sensing strategy}

The two FP-WFS algorithms implemented in METIS focus respectively on non-common path pointing errors in vortex coronagraphy modes, and on higher order non-common path aberrations (NCPA) for all HCI modes. The first one, referred to as Quadrant Analysis of Coronagraphic Images for Tip-tilt Sensing (QACITS\cite{Huby15}), uses focal-plane images obtained with the vortex coronagraph to infer the pointing offset between the target star and the center of the vortex phase ramp. It was demonstrated on sky in various infrared cameras housing a vortex coronagraph\cite{Huby17,Maire20}, and showed closed-loop control accuracy down to about $0.01\lambda/D$. The second one, reffered to as Phase Sorting Interferometry (PSI\cite{Codona13}), uses focal plane science images and simultaneous wavefront sensor (WFS) telemetry to measure the NCPA phase. The principle is based on the standard interferometry technique using a phase-shifted reference beam to probe an unknown subject beam, the key difference being that the reference beam here is random, instead of being controlled. Indeed, in our context, the reference beam (or probes) are the residual AO speckles, which are beyond our control but for which we have a known (albeit approximative) phase measurement provided by the WFS telemetry.

These two algorithms were initially designed to compensate for slowly varying pointing errors and NCPA, which can be due to differential chromatic beam wander on the METIS optics, or to any source of instability within the instrument (thermo-mechanical drifts, vibrations, etc.). Operating these two algorithms at a higher repetition frequency would however make them appropriate to sense, and correct for a significant part of WV seeing, provided that the associated wavefront variations are not too rapid. This is what we explore in the next section.

    \subsection{Potential performance gain with QACITS and PSI}

In order to investigate at which repetition frequency the QACITS and PSI control loops should be operated to provide a significant gain in terms of HCI performance in presence of WV seeing, we perform noiseless integral control simulations of the first 100 Zernike modes (modes 2--3 with QACITS, modes 4--100 with PSI) at various loop repetition frequencies, and use the resulting Zernike time series as an input to our end-to-end simulations (on top of SCAO residuals). The analysis presented here assumes that NCPA can be unbiasedly sensed by QACITS and PSI, and that sensing is done in an ``instantaneous'' (non-iterative) way. This may not be representative of PSI at high loop speed, since PSI is expected to have an intrinsic bandwidth of no more than 1~Hz due to its iterative nature. The simulations presented here should therefore be considered as upper limits to what can actually be done in terms of integral closed-loop control. 

\begin{figure}[t]
\begin{center}
\begin{tabular}{c} %% tabular useful for creating an array of images 
\includegraphics[width=\textwidth]{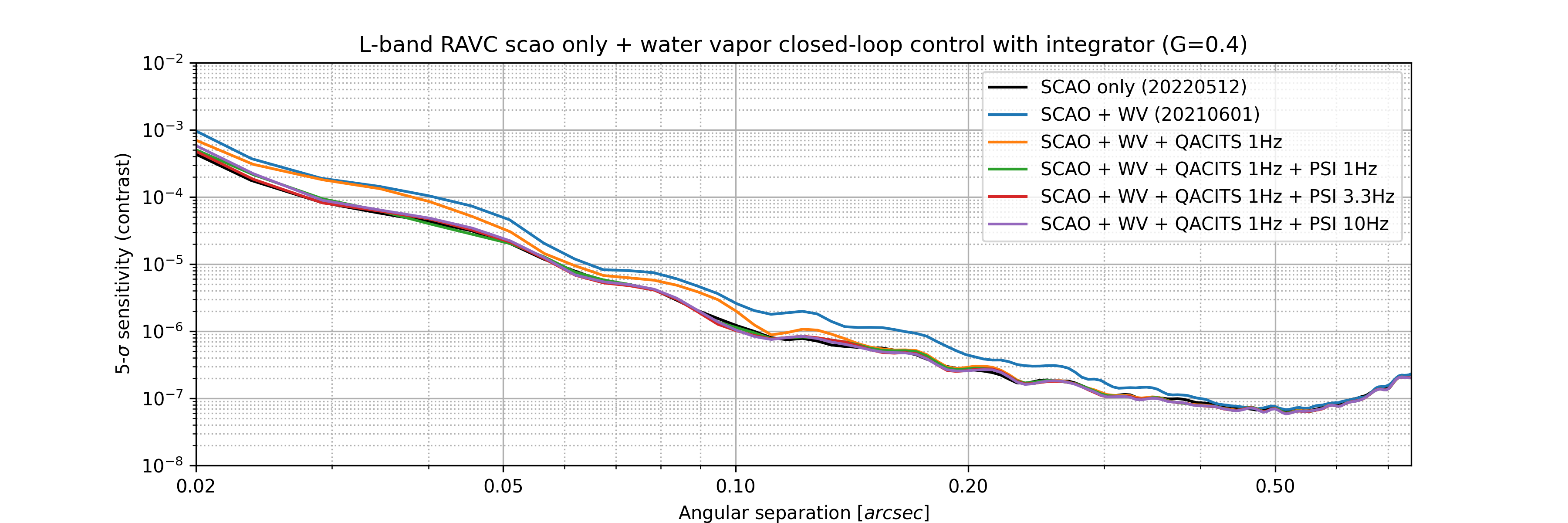} \\
\includegraphics[width=\textwidth]{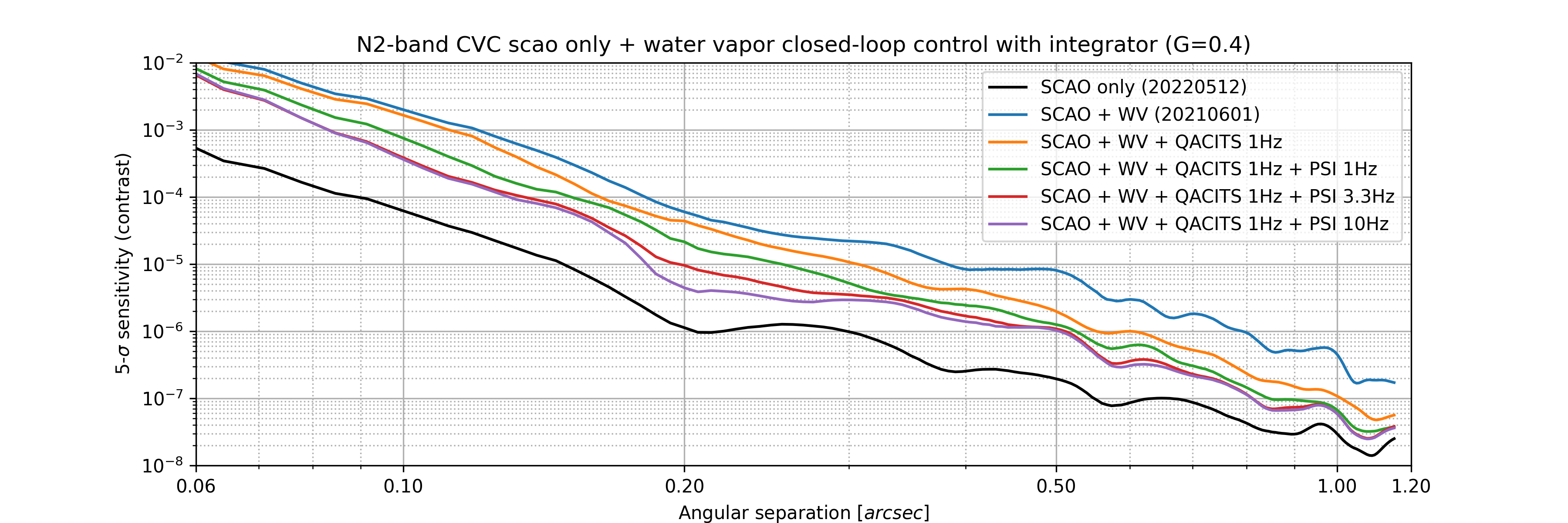}
\end{tabular}
\end{center}
\caption[example]{ \label{fig:perf_control} 
Same as Fig.~\ref{fig:perf_kolmo} in the presence of QACITS and/or PSI noiseless integral closed-loop control, at various loop repetition frequencies. An integral gain of 0.4 is used in all simulations.}
\end{figure}

The potential performance gain is illustrated in Fig.~\ref{fig:perf_control}, where we demonstrate that pointing control at 1~Hz brings the performance down to the same level as if pointing errors were completely removed (this can be seen by comparing with Fig.~\ref{fig:perf_kolmo}). For PSI, we tested repetition frequencies from 1 to 10 Hz for the control loop, showing that operating PSI above 1~Hz would be an asset. However, the iterative nature of PSI will most probably prevent effective closed-loop control above 1~Hz. Preliminary analysis of the PSI dynamical behavior suggests that a 1~Hz effective repetition frequency is within reach. With this loop repetition frequency, the L-band performance can in principle be brought back very close to the SCAO-limited performance, while the N-band performance is still degraded by about an order of magnitude compared to the SCAO-limited performance.

The simulations presented here assume noiseless sensing with QACITS and PSI. While this is unrealistic, a dedicated analysis of the noise in each of these two FP-WFS algorithm was performed as part of the METIS Phase C, and suggest that sensor noise will not be an issue for the relatively bright targets used in the HCI, up to magnitude $L\simeq 7$ for PSI and up to $L\simeq 10$ for QACITS. Noise at N band is more of an issue, as QACITS and PSI are expected to operate with a low enough noise level at 1~Hz only up to magnitude $N \simeq 2$ and $N \simeq 0$, respectively. A more concerning issue, however, is the fact that at N band, the SCAO WFS measurements used to characterize the atmospheric probes in PSI are not representative any more of the actual wavefront variability, due to the strong contribution of WV seeing even at temporal frequencies of 10~Hz and more associated with the temporal sampling from the imaging camera (typical integration time of about 100~msec). Checking whether PSI can still work under these conditions is work in progress, and may require some adaptations to the original algorithm, or even a complete change of FP-WFS framework, using e.g.\ machine learning techniques\cite{Orban21}.

\acknowledgments % equivalent to \section*{ACKNOWLEDGMENTS}       
 
This project has received funding from the European Research Council (ERC) under the European Union's Horizon 2020 research and innovation programme (grant agreement No 819155), and from the Wallonia-Brussels Federation (grant for Concerted Research Actions).  

% References
\bibliography{report} % bibliography data in report.bib
\bibliographystyle{spiebib} % makes bibtex use spiebib.bst

\end{document}